\documentclass{icrc2009}

\usepackage{graphicx}   % for including figures
\usepackage[font=footnotesize]{subfig} % subfig.sty for a double column floating figure using two subfigures
\usepackage{fixltx2e}
\usepackage{url}

\newcommand{\shorttitle}[1]%
{\markboth{Proceedings of the 31\MakeLowercase{$^{st}$} ICRC, {\L}\'{o}d\'{z} 2009}{#1} }
\begin{document}
\title{VERITAS Observation of Gamma-Ray Bursts}

\author{\IEEEauthorblockN{Nicola Galante\IEEEauthorrefmark{1} for the VERITAS Collaboration\IEEEauthorrefmark{2}}
                            \\
\IEEEauthorblockA{\IEEEauthorrefmark{1}Harvard-Smithsonian Center for Astrophysics, USA, ngalante@cfa.harvard.edu\\
\IEEEauthorrefmark{2}see R.A. Ong et al (these proceedings) for a full author list}
}

% please write the preseter's name and short title (3-4 words maximum)
%    which will appear at the header of the even pages.
\shorttitle{N. Galante, VERITAS observation of GRB}
\maketitle

\begin{abstract}
During its first cycle of observations, VERITAS observed several GRBs in response to broadcast alerts from the 
Gamma-ray bursts Coordinates Network (GCN). The GRBs were followed up and observed thereafter 
with typical delays of 2 to 4 minutes from the beginning of the burst and of 92 s in the best case, 
searching for a very high energy (VHE) component above 100 GeV. 
The aim of the search for VHE emission from GRBs is to understand the behavior, composition and 
dynamics of the most accelerated particles in the bursts, as well as to get a better overall understanding 
of the model of GRBs. We report on the results from two years of observations by VERITAS.
  \end{abstract}

 
\section{Introduction}
The study of $\gamma$-ray bursts at different wavelength has revealed many features that
suggest a connection between GRBs and some specific catastrophic events, particularly
for what concerns the class of long bursts that are thought to arise fom extremely energetic
supernovae, hypernovae. The X-ray instruments currently monitoring the sky are capable
to promptly broadcast accurate coordinates in case of a trigger due to a GRB, allowing
ground-based telescopes to perform an immediate follow-up observation.

The very upper end of $\gamma$-ray bursts has not been described yet by the very
high energy (VHE: $E > 100$~GeV) $\gamma$-ray telescopes operating so far
although several attempts have been already done by the current generation of air Cherenkov
telexcopes~\cite{Albert1}~\cite{Albert2}. 
However, several hints of a VHE component have already been given by the past and present $\gamma$-ray
observatories. 
The EGRET experiment on the CGRO detected prompt $\gamma$-rays (during $T_{90}$ of the keV/MeV emission) above 
100~MeV in four bright bursts. The composite spectrum of the bursts is well fit by a power law with a hard differential 
index of 1.95 and no indication of a spectral break or cut off up to an energy of 10~GeV~\cite{dingus2001}. 
For GRB~940217, the high energy emission continued for 90~minutes 
after the start of the burst and included an 18-GeV photon about 80~minutes after the burst onset~\cite{hurley1994}. Not only is there 
no indication of a high-energy cutoff in the spectrum of this burst, but both the prompt and delayed high-energy spectra 
show some indication of hardening above about 1~GeV. In another case, a high energy emission component, with a 
duration of 211~s, compared to the 77~s $T_{90}$ measured by BATSE, was found extending up to at least 
200~MeV in GRB~941017~\cite{gonzales2003}. This component is quite hard, with a differential power law index about 1 and with the peak 
of the $E^2 \, \mathrm{d}N/\mathrm{d}E$ distribution above the 200~MeV limit of the instrument. 
A similar component has been identiÞed in GRB~980923~\cite{gonzales2004}. The Fermi-LAT instrument recently
detected high energy emission from GRB~080916C 
continuing for at least 1300~s after the burst trigger and extending above 1~GeV.
The detection of a 13~GeV photon by the Fermi-LAT from GRB~080916C~\cite{Abdo2009}
at a redshift of $4.35\pm0.15$~\cite{Greiner2009}
shows that photons with at least 70~GeV can be produced by GRBs.
In addition, Swift-XRT detected a delayed rebrightening of the X-ray afterglow in many bursts. This would suggest the possibility
of correlated $\gamma$-ray emission originating from IC scattered photons~\cite{Wang2006}~\cite{Wang2007}.

\begin{table*}[!th]
\caption{Observed GRBs during the period between January 2008 and April 2009 after the selection
criteria explained in the text. The seven columns represent: the GRB name; the $T_{90}$;
the total delay of observation from $T_0$, that includes the delay of the GCN alert, the reaction time of the operators
and the slewing time; the observation time interval; the VERITAS energy threshold for the observation;
the flux upper limit above the energy threshold; the redshift.}\label{tab}
\centering
\begin{tabular}{lcccccc}
\hline
\hline
GRB & $T_{90}$ & $\Delta T$ & Time Interval & $E_\mathrm{th}$ & Upper Limit & redshift \\
{} & [s] & [s] & UT & [GeV]& [cm$^{-2}$ s$^{-1}$]& \\
\hline
GRB~080310 & 365 & 282 & 08:42:40 -- 12:33:40 & 350 & $4.48\times 10^{-12}$ &2.43\\
\hline
GRB~080330 & 61 & 156 & 03:43:52 -- 04:03:54 & 250 & $2.31\times 10^{-11}$ & 1.51\\
\hline
GRB~080604 & 82 & 281 & 07:31:42 -- 09:38:21 & 350 & $6.33\times 10^{-12}$ & 1.42 \\
\hline
GRB~080607 & 79 & 184 & 06:10:31 -- 06:30:33 & 450 & $1.32\times 10^{-11}$ & 3.04 \\
\hline
GRB~081024A & 1.8 & 150 & 05:55:38 -- 09:07:49 & 250 & $6.81\times 10^{-12}$ & -- \\
\hline
\end{tabular}
\label{default}
\end{table*}%

 \section{VERITAS Observations}
 
 VERITAS (the Very Energetic Radiation Imaging Telescope Array System) is an array of four 12~m aperture imaging 
atmospheric Cherenkov telescopes (IACT)~\cite{{weeks2002},{holder2006}}, 
operating at the Whipple Observatory in southern Arizona. By indirect 
detection of showers produced by $\gamma$-rays interacting in the atmosphere, VERITAS can achieve a very large
effective area, e.g. 20000 m$^2$ at 300~GeV. The showers are reconstructed from the images of the Cherenkov light they 
produce, as recorded in the $3.5^\circ$ field of view cameras of at least two of the four telescopes. At
zenith, the energy threshold for 
spectral reconstruction is about 150~GeV, and the energy resolution for individual gamma rays is 15-20\%. The 68\% 
containment radius for reconstructed $\gamma$-ray directions decreases with increasing energy, ranging from 
$0.1^\circ$ to $0.2^\circ$ . VERITAS can detect a $\gamma$-ray source with 5\% of the Crab Nebula flux in 2.5~hours. 
GRB observations during the first three hours after a burst are preauthorized by the VERITAS Time Allocation 
Committee as the highest priority observations. The VERITAS control computers monitor a socket connection to the 
GCN\footnote{see \texttt{http://gcn.gsfc.nasa.gov} for the details}. 
In the event of a GRB, an audible alarm sounds and a dialog box pops up guiding the operators to retarget 
the telescopes. The telescope slew speed of $1^\circ$~s$^{-1}$ , independently in elevation and azimuth, is the limiting delay to the 
beginning of observations for most bursts. 

VERITAS is operating with the full array system of four telescopes since September 2007.
From that date,  data on the incoming GCN
alerts with all four telescopes operating have been collected. Bursts with an uncertainty on the coordinates
larger than $5^\circ$ have been excluded.
In the period from September 2007 to April 2009, data on 15 GRBs have been recorded during dark night conditions
(no moon). Data recorded under non-optimal technical conditions have been omitted.
Work is now underway to characterize the 
low elevation and moonlight performance of VERITAS, and results for these bursts should become available in the next
future. In this paper, we present results for the 5 bursts having data at high elevation, above $\sim 45^\circ$. 
 \section{Data Analysis}
 
The data have been analyzed using the VERITAS standard VEGAS analysis package, with background rejection 
cuts optimized on a Crab-like spectrum. The direction and 
impact point of each shower is reconstructed from the positions and orientations of the shower images in the cameras. 
Candidate $\gamma$-ray showers are required to have images with length and width consistent with expectations from 
simulated showers of the corresponding size and distance from each telescope, allowing a large fraction of the 
background proton showers to be rejected. Proton showers typically have longer and wider images.

Table~\ref{tab} shows the result of the analysis performed on the five selected GRBs.
For each of the 5 bursts, there is no signiÞcant excess at the nominal position of the burst and the distribution of event 
excesses in the field of view is consistent with a Gaussian distribution expected from background fluctuations.
Integral upper limits on the photon flux are calculated at the 99\% 
conÞdence level using the method described by~\cite{Rolke} assuming a Poissonian background and a 20\%
uncertainty on the efficiency.
All upper limits are calculated above the energy threshold of the data set recorded.

\section{Conclusions}

Between Winter 2008 and Spring 2009 VERITAS could react and follow up 13 GRB events, out of which only 5
fell into the requirements for a reliable standard analysis. Observations were performed with a typical delay of
1--4 minutes since the burst $T_0$. No positive detection has been recorded
(as expected given the high redshift values of the bursts) and upper limits
above the energy threshold have been obtained.
Work to fully characterize the telescope array performances at large zenith angle and during moon time
is in progress.

\section*{aknowledgments}
 
This research is supported by grants from the US Department of  
Energy, the US National Science Foundation, and the Smithsonian  
Institution, by NSERC in Canada, by Science Foundation Ireland, and  
by STFC in the UK. We acknowledge the excellent work of the technical  
support staff at the FLWO and the collaborating institutions in the  
construction and operation of the instrument.

\end{document}